\documentclass[conference]{IEEEtran}

\usepackage{cite}
\usepackage{amsmath,amssymb,amsfonts}
\usepackage{graphicx}
\usepackage{xcolor}
\usepackage[whole]{bxcjkjatype}
\usepackage{xurl}
\usepackage{booktabs}
\usepackage{colortbl}
\usepackage{eso-pic}

\AddToShipoutPictureBG*{
  \AtPageUpperLeft{
    \raisebox{-1.2cm}{\hspace{2cm}\fbox{\parbox{0.95\textwidth}{\small This paper was originally presented at IEEE CCNC 2025. An extended version of this work has been published in IEEE Access.\\
    For citation, please refer to the IEEE Access version: D.~Chiba, H.~Nakano, and T.~Koide, ``DomainLynx: Advancing LLM Techniques for Robust Domain Squatting Detection,'' IEEE Access, 2025, https://doi.org/10.1109/ACCESS.2025.3542036.}}}
  }
  \AtPageLowerLeft{
    \raisebox{1.2cm}{\hspace{2cm}\parbox{0.95\textwidth}{\footnotesize © 2025 IEEE. Personal use of this material is permitted. Permission from IEEE must be obtained for all other uses, in any current or future media, including reprinting/republishing this material for advertising or promotional purposes, creating new collective works, for resale or redistribution to servers or lists, or reuse of any copyrighted component of this work in other works.}}
  }
}

\begin{document}

\title{DomainLynx: Leveraging Large Language Models for Enhanced Domain Squatting Detection}

\author{
\IEEEauthorblockN{Daiki Chiba, Hiroki Nakano, and Takashi Koide}
\IEEEauthorblockA{NTT Security Holdings Corporation \& NTT Corporation, Tokyo, Japan \\ Email: daiki.chiba@ieee.org}
}

\maketitle

\begin{abstract}
Domain squatting poses a significant threat to Internet security, with attackers employing increasingly sophisticated techniques. This study introduces DomainLynx, an innovative compound AI system leveraging Large Language Models (LLMs) for enhanced domain squatting detection. Unlike existing methods focusing on predefined patterns for top-ranked domains, DomainLynx excels in identifying novel squatting techniques and protecting less prominent brands. The system's architecture integrates advanced data processing, intelligent domain pairing, and LLM-powered threat assessment. Crucially, DomainLynx incorporates specialized components that mitigate LLM hallucinations, ensuring reliable and context-aware detection. This approach enables efficient analysis of vast security data from diverse sources, including Certificate Transparency logs, Passive DNS records, and zone files. Evaluated on a curated dataset of 1,649 squatting domains, DomainLynx achieved 94.7\% accuracy using Llama-3-70B. In a month-long real-world test, it detected 34,359 squatting domains from 2.09 million new domains, outperforming baseline methods by 2.5 times. This research advances Internet security by providing a versatile, accurate, and adaptable tool for combating evolving domain squatting threats. DomainLynx's approach paves the way for more robust, AI-driven cybersecurity solutions, enhancing protection for a broader range of online entities and contributing to a safer digital ecosystem.
\end{abstract}


\section{Introduction}
\label{sec:introduction}
Domain squatting, a significant cybersecurity threat, involves malicious actors registering domain names that closely resemble those of popular brands or services to deceive users. This practice poses considerable risks, potentially leading to identity theft, financial fraud, or the distribution of malware. Recent studies have shed light on the prevalence and sophistication of domain squatting techniques. Tian et al. conducted a comprehensive analysis of squatting phishing domains, identifying domains potentially impersonating 702 popular brands~\cite{DBLP:conf/imc/TianJ0Y018}. Their research highlighted the effectiveness of these deceptive tactics in evading detection, with more than 90\% of phishing pages successfully avoiding blacklisting for at least a month. Similarly, Zeng et al. explored various types of domain squatting, including typo-squatting, bit-squatting, and combo-squatting, revealing that while typo-squatting accounts for the majority of squatting domains, combo-squatting attracts more traffic~\cite{DBLP:conf/icc/ZengZ0CW19}.

While existing research has made significant strides in understanding domain squatting, several critical gaps remain. Current detection methods primarily focus on predefined squatting types for well-known brands or services. This approach, however, falls short when confronted with novel squatting techniques or domains impersonating less prominent brands. Quinkert et al. highlighted the prevalence of domain impersonation in the lower DNS hierarchy, a technique that often escapes traditional detection methods~\cite{DBLP:conf/dimva/QuinkertTH21}. Furthermore, the rapid evolution of squatting tactics poses a continuous challenge to existing detection systems, which may struggle to adapt quickly enough to new patterns. Addressing these gaps is crucial for developing more robust and comprehensive protection against domain squatting threats. Enhancing detection capabilities to cover a broader range of domains and squatting techniques could significantly reduce the success rate of these deceptive practices, thereby improving overall cybersecurity for Internet users and businesses alike.

To address these critical gaps, this study introduces DomainLynx, a novel compound AI system for domain squatting detection that leverages Large Language Models (LLMs). DomainLynx aims to overcome the limitations of traditional methods by offering a more versatile and adaptive approach to threat detection. The primary objectives of this research are threefold: (1) to demonstrate the effectiveness of LLMs in identifying a broader range of domain squatting techniques, including previously unknown or emerging patterns; (2) to evaluate the system's performance in detecting squatting attempts on both high-profile and less prominent domains; and (3) to assess the real-world applicability of DomainLynx through extensive testing on live Internet data. By achieving these goals, this study seeks to contribute significantly to the field of cybersecurity, offering a more robust defense against the evolving threat of domain squatting and enhancing the overall safety of Internet users and organizations. The development of DomainLynx represents a significant advancement in cybersecurity applications, demonstrating how AI systems can be effectively employed to process vast amounts of security log data, potentially transforming practices in detecting and mitigating online deception.

\section{Background and Related Work}
\label{sec:background}

\subsection{Types of Domain Squatting}
Domain squatting has evolved into a complex and multifaceted threat in the cybersecurity landscape~\cite{DBLP:conf/acsac/KoideFN023}. As the Internet has grown and diversified, so too have the techniques employed by malicious actors to exploit user trust and navigate around detection systems. This section examines the primary types of domain squatting techniques that have emerged over time.

\noindent\textbf{Typo-squatting.}
Typo-squatting is the most well-known and longstanding form of domain squatting~\cite{DBLP:conf/ndss/AgtenJPN15,DBLP:journals/compsec/ChibaAYHMG18}. This technique capitalizes on common typographical errors users make when entering domain names. Typo-squatting encompasses several subcategories. These include missing-dot errors, where periods between subdomains are omitted (e.g., \texttt{wwwexample[.]com}); character omission, involving the deletion of a character from the domain name (e.g., \texttt{exampl[.]com}); character permutation, where adjacent characters are transposed (e.g., \texttt{eaxmple[.]com}); character replacement, substituting characters with those adjacent on a keyboard (e.g., \texttt{exampke[.]com}); and character insertion, adding an extra character to the domain name (e.g., \texttt{examplle[.]com}).

\noindent\textbf{Homograph-squatting.}
Homograph attacks, also known as homograph-squatting, exploit the visual similarity between characters from different writing systems~\cite{DBLP:conf/dsn/LiuLLLDHZ18,DBLP:journals/jip/Sawabe0AG19}. This technique has become increasingly sophisticated with the introduction of Internationalized Domain Names (IDNs). Attackers can use visually indistinguishable characters from various Unicode blocks to create deceptive domain names. For instance, the Cyrillic character `а' (U+0430) can be used to replace the Latin `a' in a domain name, creating a visually identical but distinct URL.

\noindent\textbf{Bit-squatting.}
Bit-squatting is a more subtle and technically sophisticated form of domain squatting. As described by Nikiforakis et al.~\cite{DBLP:conf/www/NikiforakisAMDPJ13}, this technique exploits random bit errors in computer memory. When such errors occur, they can cause users to be inadvertently redirected to a malicious domain even when the correct URL is entered. For example, a single bit flip in the ASCII representation of \texttt{example[.]com} could result in \texttt{exemple[.]com}, potentially directing users to a malicious site.

\noindent\textbf{Sound-squatting.}
Sound-squatting, as introduced by Nikiforakis et al.~\cite{DBLP:conf/isw/NikiforakisBDPJ14}, targets the auditory rather than visual aspects of domain names. This technique leverages homophones—words that sound alike but are spelled differently. For instance, \texttt{eggsample[.]com} might be registered to target users of \texttt{example[.]com} who rely on text-to-speech software or mishear the domain name. This method is particularly effective in exploiting voice-based interactions with devices.

\noindent\textbf{TLD-squatting.}
With the proliferation of new generic top-level domains (gTLDs), TLD-squatting has emerged as a significant threat~\cite{DBLP:conf/imc/TianJ0Y018}. Attackers register domain names that are identical to well-known brands but under different TLDs. For example, if \texttt{example[.]com} is a legitimate site, an attacker might register \texttt{example[.]shop} or \texttt{example[.]tech}. This technique exploits user confusion about the correct TLD and the assumption that all domains associated with a brand are legitimate.

\noindent\textbf{Combo-squatting.}
Combo-squatting, as defined by Kintis et al.~\cite{DBLP:conf/ccs/KintisMLCGPNA17}, involves combining a target domain name with additional words or characters. Unlike typo-squatting, combo-squatting domains are typically not typographical variations of the target domain. For instance, \texttt{example-secure[.]com} or \texttt{myexample[.]com} could be combo-squatting domains targeting \texttt{example.com}. This technique is often employed in phishing attacks and malware distribution campaigns.

\noindent\textbf{Level-squatting.}
Level-squatting is a technique where attackers embed the target brand name within a subdomain of a malicious domain~\cite{DBLP:conf/securecomm/DuYLDHLYLSLGZL19}. For example, \texttt{example.com.domain[.]example} uses \texttt{example[.]com} as a subdomain to deceive users. This method is particularly effective on mobile devices where the full domain name may not be visible in the address bar.

\noindent\textbf{Hybrid-squatting.}
Hybrid-squatting is a new concept introduced in this study. This technique combines multiple squatting methods to create more sophisticated and harder-to-detect malicious domains. By leveraging various techniques such as typo-squatting, homograph-squatting, combo-squatting, and level-squatting in combination, attackers can generate numerous variations of legitimate domain names that are challenging for traditional detection systems to identify.

For instance, a domain such as \texttt{exarnple-secure.domain[.]example} targets \texttt{example[.]com} by combining homograph-squatting (\texttt{exarnple}) with combo-squatting (\texttt{-secure}), level-squatting (subdomain), and using a different TLD (\texttt{.example}). This combination makes the squatted domain appear legitimate while employing multiple squatting techniques to increase the chances of deceiving users.

\begin{table*}[!t]
    \caption{Overview of Existing Domain Squatting Detection Methods}
    \label{tab:squatting_detection}
    \centering
    \tabcolsep=1.4mm
    {\renewcommand\arraystretch{1.1}
    \begin{tabular}{lll | llllllll | l}
    \toprule
     &  &  & \multicolumn{8}{l|}{Detected Squatting Domain Type}  &  \\
    System & Venue & Year & Typo & Homo & Bit & Sound & TLD & Combo & Level  &Hybrid& \# Reference Legitimate Domains \\
    \midrule
    \rowcolor{gray!10}Nikiforakis et al.~\cite{DBLP:conf/www/NikiforakisAMDPJ13} & WWW & 2013 & - & - & \checkmark & - & - & - & -  &-& 500 from Alexa \\
    AutoSS~\cite{DBLP:conf/isw/NikiforakisBDPJ14} & ISC & 2014 & - & - & - & \checkmark & - & - & -  &-& 10,000 from Alexa \\
    \rowcolor{gray!10}Agten et al.~\cite{DBLP:conf/ndss/AgtenJPN15} & NDSS & 2015 & \checkmark & - & - & - & - & - & -  &-& 500 from Alexa \\
    Kintis et al.~\cite{DBLP:conf/ccs/KintisMLCGPNA17} & CCS & 2017 & - & - & - & - & - & \checkmark & -  &-& 246 from Alexa \\
    \rowcolor{gray!10}Liu et al.~\cite{DBLP:conf/dsn/LiuLLLDHZ18} & DSN & 2018 & - & \checkmark & - & - & - & - & -  &-& 1,000 from Alexa \\
    SquatPhish~\cite{DBLP:conf/imc/TianJ0Y018} & IMC & 2018 & \checkmark & \checkmark & \checkmark & - & \checkmark & \checkmark & -  &-& 702 from Alexa \\
    \rowcolor{gray!10}DomainScouter~\cite{DBLP:conf/raid/0001HKSGA19} & RAID & 2019 & - & \checkmark & - & - & - & \checkmark & -  &-& 2,310 from Alexa, Umbrella, Majestic \\
    ShamFinder~\cite{DBLP:conf/imc/Suzuki0YMG19} & IMC & 2019 & - & \checkmark & - & - & - & - & -  &-& 10,000 from Alexa \\
    \rowcolor{gray!10}LDS~\cite{DBLP:conf/securecomm/DuYLDHLYLSLGZL19} & SecureComm & 2019 & - & - & - & - & - & - & \checkmark  &-& 10,000 from Alexa \\
    Quinkert et al.~\cite{DBLP:conf/cns/QuinkertLRKH19} & CNS & 2019 & - & \checkmark & - & - & - & - & -  & -& 10,000 from Majestic \\
    \rowcolor{gray!10}Zeng et al.~\cite{DBLP:conf/icc/ZengZ0CW19} & ICC & 2019 & \checkmark & \checkmark & \checkmark & \checkmark & - & \checkmark & -  & - & 786 from Alexa \\
    Quinkert et al.~\cite{DBLP:conf/dimva/QuinkertTH21} & DIMVA & 2021 & \checkmark & \checkmark & - & - & \checkmark & \checkmark & \checkmark  & - & 1,012 from Tranco \\
    \midrule
    \rowcolor{yellow!25}DomainLynx & - & 2024 & \checkmark & \checkmark & \checkmark & \checkmark & \checkmark & \checkmark & \checkmark  &\checkmark  & LLM's Knowledge + Dynamic References\\
    \bottomrule
    \end{tabular}
    }
\end{table*}

\subsection{Existing Domain Squatting Detection Methods}
The field of domain squatting detection has seen significant advancements over the past decade, with researchers developing various techniques to identify and mitigate different types of squatting attacks. Table~\ref{tab:squatting_detection} provides an overview of the major detection systems, their capabilities, and the datasets they use.

\noindent\textbf{Detected Squatting Domain Type.}
As shown in Table~\ref{tab:squatting_detection}, research in domain squatting detection has evolved to address a wide range of squatting techniques. Early work, such as that by Nikiforakis et al.~ \cite{DBLP:conf/www/NikiforakisAMDPJ13}, focused on specific types of squatting, in this case, Bit-squatting. Subsequent studies expanded the scope of detection.
Typo-squatting detection was prominently addressed by Agten et al.~\cite{DBLP:conf/ndss/AgtenJPN15} and later incorporated into more comprehensive systems like SquatPhish~\cite{DBLP:conf/imc/TianJ0Y018}.
Homograph-squatting received attention from researchers such as Liu et al.~\cite{DBLP:conf/dsn/LiuLLLDHZ18} and have since been included in many multi-threat detection systems.
Sound-squatting, while less common, was specifically targeted by AutoSS ~\cite{DBLP:conf/isw/NikiforakisBDPJ14} and later included in more comprehensive approaches like that of Zeng et al.~\cite{DBLP:conf/icc/ZengZ0CW19}.
Combo-squatting detection emerged as a focus area, with Kintis et al.~\cite{DBLP:conf/ccs/KintisMLCGPNA17} providing seminal work in this domain.
TLD-squatting and level squatting have gained attention in more recent years, as evidenced by the work of Quinkert et al.~\cite{DBLP:conf/dimva/QuinkertTH21}.
The trend towards more comprehensive detection systems is evident, with recent works like SquatPhish~\cite{DBLP:conf/imc/TianJ0Y018} and the study by Quinkert et al.~\cite{DBLP:conf/dimva/QuinkertTH21} addressing multiple squatting techniques simultaneously.

\noindent\textbf{Reference Legitimate Domains Used in Existing Studies.}
The choice of reference legitimate domains plays a crucial role in the development and evaluation of squatting detection systems. As illustrated in Table~\ref{tab:squatting_detection}, there is considerable variation in the datasets used across studies.
Alexa Top Sites has been the most common source, with sample sizes ranging from 500 to 10,000 domains. For instance, Nikiforakis et al.~\cite{DBLP:conf/www/NikiforakisAMDPJ13} used 500 top domains from Alexa, while AutoSS~ \cite{DBLP:conf/isw/NikiforakisBDPJ14} expanded this to 10,000 domains.
Some researchers have diversified their data sources. DomainScouter~\cite{DBLP:conf/raid/0001HKSGA19} incorporated domains from Alexa, Umbrella, and Majestic, using a total of 2,310 reference domains.
More recent studies have begun to use alternative ranking services. Quinkert et al.~\cite{DBLP:conf/cns/QuinkertLRKH19} utilized 10,000 domains from Majestic, while their later work in 2021~\cite{DBLP:conf/dimva/QuinkertTH21} employed 1,012 domains from the Tranco list~\cite{DBLP:conf/ndss/PochatGTKJ19}, a newer and more stable domain ranking.

The variation in the number and sources of reference domains across studies highlights the challenge of creating a comprehensive and representative dataset for squatting detection. This diversity also underscores the need for detection methods that can generalize across different types of domains, from highly popular websites to less prominent but equally vulnerable targets.

\section{Proposed System: DomainLynx}
\label{sec:proposed-system}
DomainLynx represents a new compound AI system designed to revolutionize domain squatting detection. By augmenting Large Language Models (LLMs) with specialized components for data processing and validation, DomainLynx demonstrates the potential of AI systems to efficiently analyze vast amounts of security log data, specifically domain names in this context. This approach allows for the identification of malicious activities that might otherwise go undetected in the vast sea of Internet data.

\begin{figure*}[!t]
    \centering
    \includegraphics[width=1.0\linewidth]{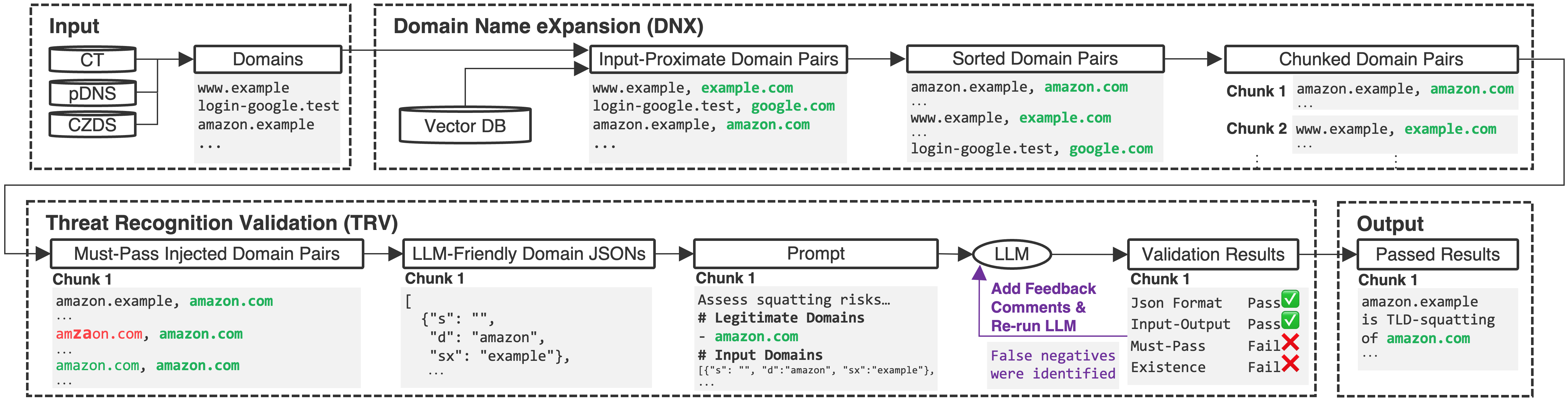}
    \caption{DomainLynx System Architecture Overview}
    \label{fig:system}
\end{figure*}

\subsection{System Architecture Overview}
DomainLynx's architecture is designed to overcome the limitations of traditional detection methods, especially in recognizing novel squatting techniques and safeguarding less prominent domains. The system comprises four primary components as shown in Figure~\ref{fig:system}, each serving a crucial role in the detection process.

First, the Input Data Processing component ingests domain data from diverse sources. Second, the Domain Name eXpansion (DNX) identifies and pairs potentially squatted domains with legitimate counterparts using vector database technology. Third, the Threat Recognition Validation (TRV) employs LLMs to assess squatting risk, incorporating advanced techniques to ensure accuracy. Finally, the Output Generation component produces a list of high-risk domains for further action.

DomainLynx's workflow progresses sequentially through these components, enabling efficient processing of vast numbers of domains while maintaining high detection accuracy. This architecture allows the system to adapt to emerging threats in the dynamic cybersecurity landscape.

\subsection{Input Data Sources}
DomainLynx ingests domain data from three primary sources to ensure comprehensive coverage of potential squatting attempts.

\noindent\textbf{Certificate Transparency (CT) Logs.}
CT logs provide a record of all SSL/TLS certificates issued by publicly trusted Certificate Authorities. These logs are valuable for discovering newly registered domains and subdomains that have obtained SSL/TLS certificates, indicating potential active use.

\noindent\textbf{Passive DNS (pDNS) Records.}
Passive DNS data is collected by monitoring DNS query responses. This source offers insights into domain resolution patterns and helps identify domains that are actively in use. pDNS data is particularly useful for detecting subdomains and identifying changes in domain resolution patterns that may indicate squatting activity.

\noindent\textbf{Centralized Zone Data Service (CZDS) Files.}
CZDS files contain authoritative information about registered domains within specific top-level domains (TLDs). This data provides a comprehensive view of domain registrations across multiple TLDs, allowing DomainLynx to track new domain registrations that may be potential squatting attempts.

\subsection{Domain Name eXpansion (DNX)}
The Domain Name eXpansion (DNX) component is a crucial part of DomainLynx, designed to efficiently identify potential squatting domains by leveraging advanced vector representation and search techniques.

\noindent\textbf{Vector Database Integration.}
DNX utilizes a vector database to store and search for domain name representations. This approach allows for semantic similarity comparisons rather than simple string matching, enabling more effective identification of potential squatting attempts.

For domain name embedding, DNX employs OpenAI's text-embedding-3-small model to convert domain names into 1536-dimensional vector representations. This model was chosen for its optimal balance of performance and efficiency, offering requirements suitable for DomainLynx's high-volume processing needs.

The system pre-populates Weaviate, an open-source vector database, with embeddings of the top 100,000 popular domain names from the Tranco list. This approach provides a reference set for identifying squatting attempts on popular domains and creates a finite space for mapping the vast number of domain names, facilitating efficient sorting and chunking in later stages.

DNX implements Retrieval Augmented Generation (RAG) to enhance the domain pairing process. For each input domain, the system generates a query embedding using the text-embedding-3-small model, performs a vector search to find the most similar domain names in the database.

\noindent\textbf{Input-Proximate Domain Pair Generation.}
Using the vector search results, DNX generates Input-Proximate Domain Pairs. These pairs consist of an input domain and its most similar counterpart from the vector database. 

\noindent\textbf{Domain Pair Sorting.}
The Input-Proximate Domain Pairs are sorted based on the Proximate Domain (legitimate domain from the vector database). This sorting strategy leverages the mapping to the Tranco top 100,000 domain space, allowing for more effective organization of input domains. By grouping input domains that are similar to the same legitimate domains together, DomainLynx enhances the efficiency of subsequent analysis and allows for better context when processing potential squatting attempts. This approach not only streamlines the analysis process but also facilitates the identification of common squatting patterns targeting specific legitimate domains, enabling more comprehensive threat detection and analysis.

\noindent\textbf{Domain Pair Chunking.}
The sorted pairs are divided into manageable chunks, typically containing 100 input FQDNs each. This chunking strategy is designed to optimize the input for the Threat Recognition Validation (TRV) component, considering the token limitations and processing capabilities of the LLM used.
The chunk size is calibrated to balance the overhead costs of prompt engineering and RAG with the need for efficient processing.
This approach leverages the asymmetric nature of LLM input-output capabilities. LLMs can process large amounts of input data in a single query but are typically constrained in the length of their output. By inputting multiple domains simultaneously, DomainLynx takes advantage of this asymmetry:

\begin{itemize}
    \item Input: The system can feed a large chunk of domain pairs (e.g., 100 pairs) along with the necessary context and instructions in a single query.
    \item Output: The LLM is instructed to provide analysis only for domains with squatting risks. Since squatting attempts typically represent a small fraction of input domains, this results in a significantly reduced output volume (e.g., 2-3 domains out of 100).
\end{itemize}

This approach significantly enhances efficiency by maximizing input processing while minimizing output generation. It leverages the fact that legitimate domains, which form the majority of inputs, do not require detailed output, thereby optimizing token usage and processing time.

By employing this sorting and chunking strategy, DomainLynx optimizes its use of LLM resources, enabling the system to process a large number of domains efficiently while maintaining the context necessary for accurate squatting detection. This approach not only enhances the system's scalability but also ensures that the subsequent TRV phase can leverage the grouped context of similar domains for more nuanced and effective threat analysis.

\begin{table*}[!t]
    \caption{Prompt for LLM-based Squatting Detection}
    \label{table:prompt_structure}
    \centering
    {\renewcommand\arraystretch{1.15}
    \begin{tabular}{lp{16cm}}
    \toprule
    Tactics & Prompt \\
    \midrule
     & \textbf{\# Task Description} \\
    Persona & You are a security analyst specialized in identifying domain squatting. Analyze a list of domain names to assess potential squatting risks.\\
    Examples & For example, \texttt{amazon.com.example.com} would be divided into \{\texttt{s}: \texttt{amazon.com}, \texttt{d}: \texttt{example}, \texttt{sx}: \texttt{com}\} where \texttt{s} is subdomain, \texttt{d} is domain, and \texttt{sx} is suffix.\\
    Steps & Using knowledge of legitimate domains, generate a JSON report detailing domains with potential risks based on squatting criteria.\\
    \midrule
     & \textbf{\# Analysis Criteria} \\
    Details & Apply the following criteria to each domain to identify squattings: \\
     & 1. \textbf{Typo-squatting}: Misspellings of domains based on keyboard proximity.  \\
    Examples & - Example: \{\texttt{d}: \texttt{faecbook}, \texttt{sx}: \texttt{com}\} targeting \texttt{facebook.com}. \\
     & 2. \textbf{Combo-squatting}: ...  \\
     & ... \\
     & 8. \textbf{Hybrid-squatting}: Combining multiple squatting techniques. \\
     & - Example: \{\texttt{s}: \texttt{amazeon-auth}, \texttt{d}: \texttt{example}, \texttt{sx}: \texttt{top}\} targeting \texttt{amazon.com}. \\
    \midrule
     & \textbf{\# Output Specification} \\
    Steps & Generate a JSON report for domains showing squatting risks, including: \texttt{s}, \texttt{d}, \texttt{sx}, \texttt{type} (most likely squatting type), \texttt{l} (targeted legitimate domain) ... \\
    \midrule
     & \textbf{\# Additional Legitimate Domains} \\
    References & Use the following legitimate domains in addition to your knowledge: \\
     & - \texttt{amazon.com} \\
     & - ... \\
    \midrule
     & \textbf{\# Input Domains} \\
    Delimiters & [\{\texttt{s}: \texttt{www}, \texttt{d}: \texttt{amazon}, \texttt{sx}: \texttt{com}\}, \{\texttt{s}: \texttt{login}, \texttt{d}: \texttt{amaz0n}, \texttt{sx}: \texttt{com}\}, ...]\\
    \bottomrule
    \end{tabular}
    }
\end{table*}

\subsection{Threat Recognition Validation (TRV)}
\label{sec:trv}
The Threat Recognition Validation (TRV) component forms the core of DomainLynx, leveraging Large Language Models (LLMs) to assess the squatting risk of domain pairs. This choice of LLMs is driven by their unparalleled ability to adapt to evolving squatting techniques and interpret nuanced linguistic patterns, offering a more flexible and comprehensive approach than traditional rule-based systems. TRV incorporates sophisticated techniques to ensure accurate and reliable results while mitigating potential LLM hallucinations.

\noindent\textbf{Must-Pass Domain Injection.}
To enhance the reliability of LLM outputs and detect potential hallucinations, TRV implements a must-pass domain injection mechanism. This process involves three key steps: selection, placement, and validation.

In the selection phase, TRV injects a set of unambiguous domains that any competent system should classify with 100\% accuracy. These include legitimate domains (e.g., \texttt{amazon[.]com}) and blatant squatting domains (e.g., \texttt{amzaon[.]com}), which we term ``must-pass'' domains. For placement, these domains are strategically inserted into each input chunk. For instance, in a chunk of 100 domains, four must-pass domains are placed at the 26th, 51st, 76th, and 101st positions. This strategy helps mitigate the ``lost in the middle'' phenomenon~\cite{DBLP:journals/tacl/LiuLHPBPL24}, where LLMs may overlook information in the center of large inputs.

The validation step involves verifying the LLM's classification of these must-pass domains. Any misclassification triggers a re-evaluation of the entire chunk, ensuring that only reliable outputs are utilized. This approach serves as a critical quality control mechanism, significantly enhancing the robustness and reliability of DomainLynx's squatting detection capabilities.

\noindent\textbf{LLM-Friendly Domain JSON Conversion.}
To facilitate accurate domain structure understanding by the LLM, input domains are converted into a structured JSON format. This format is designed as {\texttt{s}: \texttt{subdomain}, \texttt{d}: \texttt{domain}, \texttt{sx}: \texttt{suffix}}, where \texttt{s} represents the subdomain, \texttt{d} represents the domain (registrable portion), and \texttt{sx} represents the public suffix. For example, the domain \texttt{www.example.co[.]jp} would be converted to {\texttt{s}: \texttt{www}, \texttt{d}: \texttt{example}, \texttt{sx}: \texttt{co.jp}}.

This structured format offers several advantages. First, it clearly delineates the domain hierarchy, enabling the LLM to accurately parse and analyze domain structures. Second, by separating the suffix from the main domain, it enables better recognition of TLD-squatting and level-squatting attempts. This approach significantly enhances the LLM's ability to identify and categorize various types of domain squatting techniques, contributing to the overall effectiveness of the DomainLynx system.

\noindent\textbf{Prompt Engineering and LLM Integration.}
TRV employs advanced prompt engineering tactics to optimize LLM performance. Table~\ref{table:prompt_structure} in the paper provides a detailed example of the prompt structure, which encompasses several key elements designed to enhance the LLM's domain squatting detection capabilities.

The prompt begins with a clear task description, outlining the objective of analyzing domain names for potential squatting risks. This is followed by a persona setting, which establishes the LLM's role as a specialized security analyst, framing its responses within the appropriate expertise. Concrete examples are provided to illustrate how domain names are divided into their constituent parts, enhancing the LLM's understanding of the input format.

The prompt then outlines the step-by-step process the LLM should follow, including the use of legitimate domain knowledge and the generation of a JSON report based on specific squatting criteria. These criteria are clearly defined, covering various types of domain squatting such as typo-squatting and combo-squatting, each accompanied by illustrative examples.

Output specifications are detailed, including the expected format of the LLM's response and the specific fields to be included in the JSON report (\texttt{s}, \texttt{d}, \texttt{sx}, \texttt{type}, \texttt{l}). To further enhance accuracy, the prompt provides additional references in the form of known legitimate domains, supplementing the LLM's existing knowledge base.

Finally, the prompt employs clear input delimitation, marking the beginning of the input domain list. This ensures the LLM can effectively distinguish between instructions and the data to be analyzed, contributing to the overall precision of the squatting detection process.

This comprehensive prompt structure guides the LLM through the analysis process, ensuring consistent and accurate evaluations of potential squatting domains.

A key differentiator of DomainLynx, as highlighted in Table~\ref{tab:squatting_detection} is its source-independent approach to reference legitimate domains. Unlike traditional methods that rely solely on specific lists like Alexa or Tranco top domains, DomainLynx leverages two complementary sources:

\begin{itemize}
    \item Pre-existing Knowledge: The LLM's inherent knowledge of domains and brands, accumulated through its pre-training on vast amounts of Internet data.
    \item Dynamic References: Additional legitimate domains retrieved from the vector database, which may include but is not limited to the Tranco top 100k domains.
\end{itemize}

This dual approach allows DomainLynx to maintain a broader, more adaptable reference base. By combining the LLM's extensive pre-trained knowledge with dynamically sourced references, DomainLynx achieves a level of flexibility and comprehensiveness in legitimate domain recognition that is not constrained by any single source. This source-independent strategy enhances DomainLynx's ability to detect squatting attempts across a wider range of domains, including less prominent or emerging brands that might be overlooked by traditional, fixed-list approaches.

A notable feature of DomainLynx's prompt engineering is the inclusion of instructions for detecting \textit{Hybrid-squatting}, a new category of domain squatting introduced in this paper. Hybrid-squatting combines multiple squatting techniques, making it particularly challenging to detect using traditional methods. By explicitly instructing the LLM to identify this complex form of squatting, DomainLynx enhances its capability to recognize sophisticated and evolving threats.

\noindent\textbf{Validation Results Processing.}
TRV implements a rigorous four-step validation process to ensure the accuracy and reliability of LLM outputs. This comprehensive approach addresses various aspects of output quality and consistency.

The first step is JSON Format Validation. This process checks for completeness and adherence to the specified format, detecting any extraneous outputs or missing/additional keys. If the format is invalid, indicating a potential misunderstanding of instructions, the system triggers a re-run with appropriate feedback.

Next, the Input-Output Consistency Check verifies that all domains in the LLM's output correspond to the input chunk. This step is crucial for identifying potential hallucinations where the LLM might generate domains not present in the input. When inconsistencies are detected, the system provides feedback for a re-run, emphasizing the need for thorough examination of the input domains.

The third step, Must-Pass Domain Verification, confirms the correct classification of injected must-pass domains. This process is designed to detect both false negatives (missed malicious domains) and false positives (misclassified benign domains). In cases of misclassification, the system triggers a re-run with generalized feedback, encouraging a comprehensive review of all input domains without explicitly pointing out which domains were misclassified.

Finally, the Legitimate Domain Existence Verification step checks if the LLM-identified targeted legitimate domains exist in the Tranco Top 100k list. For domains not found in this list, the system actively searches using a search engine to confirm their existence. If non-existent domains are hallucinated, the results are rejected, and a re-run is initiated with appropriate feedback.

This multi-layered validation process significantly enhances the reliability of DomainLynx's output, ensuring that the system produces accurate and trustworthy results in domain squatting detection.

\subsection{Output Generation}
The Output Generation module is the final component of DomainLynx, responsible for producing a refined list of potential squatting domains based on the validated results from the Threat Recognition Validation (TRV) component.

\noindent\textbf{Iterative Validation Process.}
DomainLynx employs an iterative validation process to ensure high-quality outputs. The system repeatedly runs LLM analysis on chunks that fail validation, incorporating feedback from previous attempts. This process continues until all validation criteria are met or a predefined maximum number of attempts is reached. Crucially, only results that pass all validation checks are included in the final output. This rigorous approach ensures that the system's findings are both accurate and reliable, minimizing false positives and negatives in squatting detection.

\noindent\textbf{Output Format.}
The final output of DomainLynx consists of a structured list of potential squatting domains. Each entry in this list is comprehensive, containing three key pieces of information: the Input Domain, which is the full domain name identified as a potential squatting attempt; the Squatting Type, specifying the particular squatting technique detected; and the Targeted Legitimate Domain, indicating the legitimate domain that the squatting attempt is likely targeting.

This structured output format provides security professionals with clear, actionable information about potential domain squatting threats. By presenting the data in this manner, DomainLynx facilitates efficient response and mitigation strategies, allowing for quick identification and assessment of potential risks. The inclusion of the squatting type and targeted legitimate domain adds context to each detection, enhancing the system's utility in real-world cybersecurity applications.

\section{Evaluation}
\label{sec:evaluation}

\subsection{Experimental Setup}
To comprehensively evaluate the performance of DomainLynx, we conducted experiments using both a curated Ground Truth Dataset and a Real-time Dataset. Additionally, we implemented baseline methods based on previous studies for comparison.

\noindent\textbf{Ground Truth Dataset.}
We constructed a Ground Truth Dataset comprising 1,649 squatting domains targeting the Tranco Top 1k domains~\cite{DBLP:conf/ndss/PochatGTKJ19}. These domains were carefully selected based on predefined squatting definitions and validated using VirusTotal analysis reports. Domains flagged as malicious by at least five engines were included in the dataset. The dataset includes various squatting types: 369 typosquatting, 249 homograph, 136 bitsquatting, 33 soundsquatting, 156 TLD squatting, and 706 combination squatting domains. This diverse composition serves to evaluate DomainLynx's ability to detect various squatting techniques, particularly those targeting high-profile domains.

\noindent\textbf{Real-time Dataset.}
To evaluate DomainLynx's performance in genuine threat environments, we analyzed actual newly observed FQDNs from January 1 to 31, 2024. This month-long period captures real-world domain registration patterns and potential threats. We aggregated live data from all public CT logs, pDNS records from over 60 global DNS servers, and CZDS files from 1,430 TLDs. To focus on domains with higher threat potential, we filtered for FQDNs with both NS (Name Server) records and A (IPv4) or AAAA (IPv6) records, indicating active use. This process yielded 2,099,184 FQDNs for analysis. This extensive, real-world dataset enables a rigorous evaluation of DomainLynx against authentic and diverse threats, reflecting actual cybersecurity challenges and demonstrating the system's ability to detect sophisticated domain squatting techniques in live, dynamic online environments.

\noindent\textbf{Baseline Methods.}
To benchmark DomainLynx's performance, we implemented baseline methods based on two recent studies. From Zeng et al.\cite{DBLP:conf/icc/ZengZ0CW19}, we reimplemented their methods for detecting Typo, Combo, Homograph, and Bit squatting domains. Additionally, we incorporated approaches from Quinkert et al.\cite{DBLP:conf/dimva/QuinkertTH21} for detecting Level and TLD squatting. For fair comparison, we used the Tranco Top 1k domains as the set of legitimate brands, aligning with the scale of brand selections in these previous studies (786 and 1,014 domains, respectively).
These baseline methods represent the current state-of-the-art in domain squatting detection that do not require access to user DNS or web traffic data, making them suitable for comparison with DomainLynx in a real-time detection scenario.

\noindent\textbf{LLMs.}
This study employed three distinct Large Language Models (LLMs) for evaluation: GPT-3.5 and GPT-4o, both commercial models developed by OpenAI and accessed via their API, and Llama-3-70B, an open-source model created by Meta AI. Llama-3-70B was utilized through Groq's API, which offered the fastest inference times at the time of evaluation. These models were selected to provide a comprehensive assessment of DomainLynx's performance across different LLM architectures and capabilities.

\begin{table*}[!t]
    \caption{Detection Accuracy (\%) by Squatting Type}
    \label{table:accuracy-ground-truth}
    \centering
    {\renewcommand\arraystretch{1.15}
    \begin{tabular}{lll|rrrrrrrr}
    \toprule
    \multicolumn{3}{l|}{DomainLynx Parameters}& \multicolumn{8}{l}{Detection Results} \\
    LLM & DNX & TRV & Typo & Homo & Bit & Sound & TLD & Combo & Total & Time \\
    \midrule
    GPT-3.5 & - & - & 77.8\% & 57.4\% & 92.6\% & \textbf{100\%} & 6.4\% & 30.7\% & 49.5\% & 557 \\
    GPT-3.5 & \checkmark & - & 96.7\% & 85.1\% & 96.3\% & \textbf{100\%} & 72.4\% & 76.8\% & 84.2\% & 833 \\
    GPT-3.5 & \checkmark & \checkmark & \textbf{98.1\%} & 96.8\% & \textbf{98.5\%} & \textbf{100\%} & 88.5\% & 81.4\% & 89.9\% & 1,800 \\
    \midrule
    GPT-4o & - & - & 90.8\% & 97.6\% & 91.9\% & \textbf{100\%} & 84.6\% & 94.5\% & 93.1\% & 1,725 \\
    GPT-4o & \checkmark & - & 95.1\% & 99.2\% & 96.3\% & \textbf{100\%} & \textbf{95.5\%} & 95.3\% & 96.1\% & 1,928 \\
    GPT-4o & \checkmark & \checkmark & 94.9\% & \textbf{99.6\%} & 94.9\% & \textbf{100\%} & \textbf{95.5\%} & 95.9\% & \textbf{96.2\%} & 1,951 \\
    \midrule
    Llama-3-70B & - & - & 89.2\% & 97.6\% & 72.1\% & 97.0\% & 50.0\% & 92.4\% & 86.8\% & \textbf{184} \\
    Llama-3-70B & \checkmark & - & 89.7\% & 99.2\% & 92.6\% & \textbf{100\%} & 76.3\% & 95.2\% & 92.7\% & 211 \\
    \rowcolor{yellow!25}Llama-3-70B & \checkmark & \checkmark & 93.2\% & 98.8\% & 91.9\% & \textbf{100\%} & 85.9\% & \textbf{96.2\%} & 94.7\% & 229 \\
    \bottomrule
    \end{tabular}
    }
\end{table*}

\begin{table}[!t]
    \caption{Cost Estimate (USD) per 1M Input Domains}
    \label{table:cost}
    \centering
    {\renewcommand\arraystretch{1.1}
    \begin{tabular}{lrrr}
    \toprule
     & Input & Output & Total \\
    \midrule
    GPT-3.5 & \$25.00 & \$15.00 & \$40.00 \\
    GPT-4o & \$250.00 & \$150.00 & \$400.00 \\
    Llama-3-70B & \$29.50 & \$7.90 & \$37.40 \\
    \bottomrule
    \end{tabular}
    }
\end{table}

\subsection{Performance on Ground Truth Dataset}
We evaluated DomainLynx's performance on the Ground Truth Dataset using three LLMs. We also assessed the impact of the Domain Name eXpansion (DNX) and Threat Recognition Validation (TRV) components on detection accuracy.
Table~\ref{table:accuracy-ground-truth} presents the detection accuracy and processing time for each squatting type across different LLM configurations.

\noindent\textbf{Impact of DNX and TRV Components.}
The results demonstrate that enabling both DNX and TRV components consistently improves detection accuracy across all LLM configurations. For instance, with Llama-3-70B, the overall accuracy increased from 86.8\% (without DNX and TRV) to 94.7\% (with both DNX and TRV enabled).

\noindent\textbf{Comparison of LLM Models.}
While GPT-4o achieved the highest overall accuracy (96.2\%), Llama-3-70B with DNX and TRV enabled performed comparably well (94.7\%). Notably, Llama-3-70B outperformed GPT-3.5 in all configurations, suggesting that it offers a good balance between performance and potential cost-effectiveness.

\noindent\textbf{Processing Time.}
Llama-3-70B consistently demonstrated the fastest processing times, with the full configuration (DNX and TRV enabled) completing in just 229 seconds, significantly outperforming both GPT-3.5 and GPT-4o.

\noindent\textbf{Cost Analysis.}
We estimated the cost for processing 1 million input domains (equivalent to 50M input tokens and 10M output tokens) using API-based services for each LLM.
Llama-3-70B offers the most cost-effective solution, with a total estimated cost of \$37.40 per million domains, closely followed by GPT-3.5 at \$40.00. GPT-4o, while providing slightly higher accuracy, comes at a significantly higher cost of \$400.00 per million domains.

\noindent\textbf{Summary.}
Based on these results, we conclude that Llama-3-70B with both DNX and TRV components enabled offers the best balance of detection accuracy, processing speed, and cost-effectiveness for the DomainLynx system. This configuration achieves high accuracy across all squatting types while maintaining efficient processing times and reasonable operational costs.

\begin{table*}[!t]
    \caption{Detection Results - DomainLynx vs. Baseline}
    \label{table:detection-realtime}
    \centering
    {\renewcommand\arraystretch{1.1}
    \begin{tabular}{lrrrrrrrrr}
    \toprule
    System & \# Typo & \# Homo & \# Bit & \# Sound & \# TLD & \# Combo & \# Level & \# Hybrid & \# Total\\
    \midrule
    Baseline & 80 & 33 & 22 & 8 & 599 & 13,227 & 20 & 0 & 13,989 \\
    \rowcolor{yellow!25}DomainLynx & 5,898 & 441 & 132 & 292 & 3,177 & 23,488 & 767 & 164 & 34,359 \\
    \bottomrule
    \end{tabular}
    }
\end{table*}

\subsection{Real-world Performance Evaluation}
To assess DomainLynx's effectiveness in real-world scenarios, we conducted a one-month evaluation using newly observed FQDNs from January 1 to January 31, 2024. This evaluation aimed to test the system's performance on a large scale and compare it with baseline methods.

\noindent\textbf{Detection Results on Newly Observed FQDNs.}
During the evaluation period, DomainLynx processed 2,099,184 newly observed FQDNs. The system, configured with Llama-3-70B and both DNX and TRV components enabled, detected a total of 34,359 potential squatting domains. Table~\ref{table:detection-realtime} presents a breakdown of the detection results by squatting type, compared to the baseline methods. 

DomainLynx significantly outperformed the baseline methods, detecting 2.45 times more potential squatting domains (34,359 vs. 13,989). Notably, DomainLynx identified 164 hybrid-squatting cases, which the baseline methods were unable to detect due to their rule-based nature.

\noindent\textbf{Processing Time and Cost Efficiency.}
DomainLynx demonstrated impressive efficiency in processing the real-time dataset:

\begin{itemize}
    \item Processing Time: The system completed the analysis of 2,099,184 FQDNs in 1.5 days, which is well within the constraints of real-time operation for newly observed domains over a 31-day period.
    \item Cost: The total cost for processing the month's data using Llama-3-70B was \$57.5, indicating high cost-effectiveness for large-scale, continuous operation.
\end{itemize}

These results underscore DomainLynx's capability to handle real-world data volumes efficiently and economically.

\noindent\textbf{Validation of Detected Squatting Domains.}
To assess the accuracy of DomainLynx's detections, we employed multiple validation strategies. First, we applied input filtering by focusing on newly observed FQDNs, inherently eliminating the possibility of false positives from long-established, legitimate domains. Second, we conducted manual verification on all 34,359 detected domains more than a month after their initial detection. This process involved accessing the domains and using search engines to look for evidence of legitimate services. Notably, no clear indications of legitimate, publicly accessible services were found among the detected domains.

Lastly, we performed periodic VirusTotal analysis, scanning the detected domains using 89 different detection engines. The results revealed that 9,557 out of 34,359 domains (27.8\%) detected by DomainLynx were flagged as malicious by at least one VirusTotal engine. For comparison, 3,725 out of 13,989 domains (26.6\%) detected by the baseline methods were similarly flagged.

These validation results suggest that DomainLynx maintains a high level of precision while significantly improving recall compared to baseline methods. The similar proportion of VirusTotal-flagged domains between DomainLynx and the baseline (27.8\% vs. 26.6\%) indicates that DomainLynx's increased detection rate does not come at the cost of reduced precision.

Moreover, the fact that DomainLynx detected a substantial number of potentially malicious domains not flagged by VirusTotal underscores its ability to identify subtle or emerging threats that may evade traditional detection methods.

\noindent\textbf{Summary.}
In conclusion, the real-world evaluation demonstrates DomainLynx's superior performance in detecting a wide range of squatting domains, including complex cases like hybrid-squatting. The system's efficiency in processing large volumes of data and its cost-effectiveness make it well-suited for continuous, real-time squatting detection in practical cybersecurity applications.

\begin{table*}[!t]
    \caption{Distribution of Targeted Domain Ranking}
    \label{table:targeted-domain-ranking}
    \centering
    {\renewcommand\arraystretch{1.1}
    \begin{tabular}{lrrrrrrrrrr}
    \toprule
    Rank & Typo & Homo & Bit & Sound & TLD & Combo & Level & Hybrid & Total & \%\\
    \midrule
    \textless 1k & 1,111 & 43 & 38 & 73 & 188 & 3,221 & 218 & 38 & 4,930 & 14.4\% \\
    1k--10k & 884 & 29 & 27 & 32 & 520 & 4,571 & 165 & 33 & 6,261 & 18.2\%\\
    10k--100k & 3,242 & 87 & 54 & 149 & 2,419 & 14,726 & 326 & 86 & 21,089 & 61.4\%\\
    100k--1M & 101 & 4 & 3 & 9 & 13 & 220 & 12 & 0 & 362 & 1.1\% \\
    \textgreater 1M & 560 & 278 & 10 & 29 & 37 & 750 & 46 & 7 & 1,717 & 5.0\%\\
    \midrule
    Total & 5,898 & 441 & 132 & 292 & 3,177 & 23,488 & 767 & 164 & 34,359 & 100\%\\
    \bottomrule
    \end{tabular}
    }
\end{table*}

\subsection{Analysis of Detected Squatting Domains}
To gain deeper insights into DomainLynx's capabilities, we conducted a comprehensive analysis of the 34,359 squatting domains detected during the one-month real-world evaluation. This analysis focused on the distribution of targeted legitimate domains across popularity rankings and the system's ability to detect squatting attempts on less prominent domains.

We categorized the detected squatting domains based on the Tranco Top List ranking of their targeted legitimate domains. Table~\ref{table:targeted-domain-ranking} presents this distribution. Key findings from this analysis are as follows.

\noindent\textbf{Detection of Minor Domain Squatting.}
While 4,930 (14.4\%) of detected squatting domains targeted highly popular sites (Tranco rank \textless 1k), a significant 29,429 (85.7\%) targeted less prominent domains (Tranco rank \textgreater 1k).
Notably, 21,089 (61.4\%) of detections were associated with domains ranked between 10k and 100k, demonstrating DomainLynx's effectiveness in identifying threats to a broad range of websites beyond just the most popular ones.
The system even detected 362 (1.1\%) squatting attempts on domains ranked between 100k and 1 million, and 1,717 (5.0\%) on domains ranked beyond 1 million or not listed in the Tranco Top 1M, showcasing its ability to protect less visible but potentially vulnerable online entities.

\noindent\textbf{Squatting Type Distribution.}
Combo-squatting was the most prevalent type across all ranking categories, with 23,488 instances accounting for 68.4\% of all detections. This highlights the popularity of this technique among attackers and the importance of DomainLynx's ability to detect it effectively.
Typo-squatting was the second most common type with 5,898 instances (17.2\% of all detections), with a notable concentration of 3,242 cases in the 10k--100k range, suggesting that attackers frequently target mid-tier domains with this technique.
TLD-squatting showed a significant presence with 3,177 cases (9.3\% of all detections), particularly for domains ranked between 10k and 100k (2,419 instances), indicating a trend of exploiting alternative top-level domains for mid-popularity websites.

\noindent\textbf{Hybrid-squatting Detection.}
DomainLynx identified 164 instances (0.5\% of all detections) of Hybrid-squatting, a sophisticated technique that combines multiple squatting methods. These cases were distributed across various ranking categories, with the majority (86 instances) targeting domains ranked between 10k and 100k.

\noindent\textbf{Versatility Across Squatting Types.}
DomainLynx showed consistent performance across all squatting types for various domain popularity levels, indicating its versatility in detecting diverse squatting techniques.
The system's ability to identify less common squatting types, such as Bit-squatting (132 instances, 0.4\% of all detections) and Sound-squatting (292 instances, 0.9\% of all detections) across different ranking categories demonstrates its comprehensive approach to threat detection.

\noindent\textbf{Summary.}
This analysis reveals DomainLynx's significant advantage over traditional methods, particularly in its ability to detect squatting attempts on a wide range of domains, including less prominent ones. By leveraging LLMs and advanced processing techniques, DomainLynx overcomes the limitations of rule-based systems, offering more comprehensive protection against diverse and evolving squatting threats. This capability is crucial in the current cybersecurity landscape, where attackers increasingly target a broader spectrum of online entities beyond just the most popular websites.

\section{Discussion}
\label{sec:discussion}

\subsection{Ethical Considerations}
In the development and evaluation of DomainLynx, we prioritized data privacy and ethical use of information. Our study utilized three primary data sources: Certificate Transparency (CT) logs and Centralized Zone Data Service (CZDS) files, which are publicly available, and Passive DNS (pDNS) records from our organization's proprietary dataset. The pDNS data, collected between caching and authoritative DNS servers, contains no personally identifiable information (PII). It comprises only FQDNs and corresponding IP addresses, excluding any user-specific or traffic-identifying data. This approach allowed us to develop an effective domain squatting detection system while maintaining high standards of privacy protection and ethical data usage.

\subsection{Limitations}
While DomainLynx has demonstrated high detection performance, we acknowledge several limitations:

\noindent\textbf{LLM Dependency.}
The system's performance is heavily reliant on the quality of the Large Language Models (LLMs) used. Updates or changes to these LLMs may impact detection results, necessitating periodic system adjustments to align with LLM evolution. For instance, if an LLM's training data is updated to include more recent domain naming conventions, it might alter the system's sensitivity to certain types of domain squatting.

\noindent\textbf{Evolving Threats.}
There is a possibility that attackers may develop new techniques to evade LLM-based detection. For example, adversaries might create domain names that exploit specific weaknesses in LLM comprehension, such as using subtle semantic shifts that are difficult for current models to detect. Consequently, DomainLynx requires continuous updates and improvements to stay ahead of emerging threats in the rapidly changing landscape of cybersecurity.

\noindent\textbf{Detection Accuracy Challenges.}
Despite DomainLynx's high detection accuracy, completely eliminating false positives remains challenging. Some detected squatting domains may not be flagged by VirusTotal or show clear malicious intent upon manual verification. This limitation includes the potential for false positives, where legitimate domain names might be mistakenly flagged as squatting attempts. For example, regional variants of brand websites or intentionally similar domain names used for legitimate purposes could be incorrectly identified as threats.

These limitations underscore the need for ongoing refinement of DomainLynx and emphasize the importance of long-term evaluation in real-world scenarios. Addressing these challenges will be crucial for further enhancing the system's reliability and effectiveness in combating domain squatting threats.

\section{Conclusion}
\label{sec:conclusion}
This study introduced DomainLynx, an innovative domain squatting detection system leveraging Large Language Models (LLMs). DomainLynx addressed limitations of traditional methods by identifying a broader range of squatting techniques, including those targeting less prominent domains.

In evaluations, DomainLynx achieved 94.7\% accuracy on our Ground Truth Dataset using Llama-3-70B with Domain Name eXpansion (DNX) and Threat Recognition Validation (TRV) components. Real-world testing over one month identified 34,359 potential squatting domains from 2,099,184 newly observed FQDNs, outperforming baseline methods by 2.45 times. Notably, 85.65\% of detections targeted domains beyond the top 1,000 in popularity, demonstrating DomainLynx's effectiveness in protecting less visible online entities.

While acknowledging limitations such as LLM dependency and evolving threats, DomainLynx represents a significant advancement in squatting detection. Its ability to process large volumes of data efficiently and detect sophisticated techniques, including hybrid-squatting, positions it as a valuable cybersecurity tool.

This research contributes to the field by showcasing the potential of LLM-based approaches in cybersecurity, particularly for identifying subtle and emerging threats. As the digital landscape evolves, tools like DomainLynx will play a crucial role in safeguarding online interactions and maintaining trust in digital platforms.

\bibliographystyle{IEEEtran}
\bibliography{main}

\end{document}